\documentclass{pos}
\usepackage{multirow}
\usepackage{amsmath}

\DeclareMathAlphabet{\mathsfslbf}{OT1}{txss}{b}{sl}

\newcommand{\fb} {{\rm fb}^{-1}}

\newcommand{\lumi} {4.4~\fb}

\newcommand{\TeV}{{\rm TeV}}

\newcommand{\GeVc}{{\rm GeV/}c}
\newcommand{\GeVcsq}{{\rm GeV/}c^2}

\newcommand{\kmmBr}{0.38}
\newcommand{\kmmBrStat}{0.05}
\newcommand{\kmmBrSyst}{0.03}
\newcommand{\kstmmBr}{1.06}
\newcommand{\kstmmBrStat}{0.14}
\newcommand{\kstmmBrSyst}{0.09}
\newcommand{\phimmBr}{1.44}
\newcommand{\phimmBrStat}{0.33}
\newcommand{\phimmBrSyst}{0.46}

\newcommand{\bz}{B^0}

\newcommand{\bp}{B^+}

\newcommand{\bs}{B^0_s}
\newcommand{\bsb}{\overline{B}{}^0_s}

\newcommand {\kst}{K^{*0}}

\newcommand{\kmm}{K^+ \mu^+ \mu^-}

\newcommand{\kstmm}{\kst \mu^+ \mu^-}

\newcommand{\phimm}{\phi \mu^+ \mu^-}

\newcommand{\hmm}{h \mu^+ \mu^-}

\newcommand{\bsll}{b \to s \ell \ell}

\newcommand{\jpsi}{J/{\small \psi}}

\newcommand{\psik}{\jpsi K^+}

\newcommand{\psiphi}{\jpsi \phi}
\newcommand{\psih}{\jpsi h}

\newcommand{\bpkmm}{\bp \to \kmm}

\newcommand{\bzkstmm}{\bz \to \kstmm}

\newcommand{\bsphimm}{\bs \to \phimm}

\newcommand{\bhmm}{B \to \hmm}

\newcommand{\bppsik}{\bp \to \psik}

\newcommand{\bspsiphi}{\bs \to \psiphi}

\newcommand{\fl}{\rm F_{L}}
\newcommand{\afb}{\rm A_{FB}}

\newcommand{\cosm}{\cos \theta_\mu}
\newcommand{\cosk}{\cos \theta_K}

\newcommand{\Mmm}{M_{\mu\mu}}

\newcommand{\Bsmm}{\bs \to \mu^+\mu^-}
\newcommand{\Bdmm}{\bz \to \mu^+\mu^-}
\newcommand{\brBsmm}{{\cal B}(\Bsmm)}
\newcommand{\brBdmm}{{\cal B}(\Bdmm)}

\title{
Rare decays / $\mathsfslbf B_{\mathsfslbf s}$ CPV measurements at Tevatron
}

\ShortTitle{Rare decays / $B_s$ CPV measurements at Tevatron}

\author{\speaker{Hideki Miyake}\thanks{On behalf of the CDF and D\O\ Collaborations.}\\
        University of Tsukuba\\
        E-mail: \email{miyake@hep.px.tsukuba.ac.jp}}

\abstract{
Measurements of Flavor Changing Neutral Current (FCNC) processes (rare decays, flavor mixing) play a key role to pursue new physics beyond the Standard Model.
We present recent analysis results about some FCNC transitions performed by CDF and D0 collaborations, including the first observation of the $\bsphimm$ mode, the forward-backward asymmetry measurement in $B\to K^{(*)}\mu^+\mu^-$, and updated measurements of $\bs(\bz)\to \mu^+\mu^-$, 
using data corresponding to integrated luminosities from $3.7\fb$ to $5\fb$.
We also show the CDF/D0 combined measurement of the $\bs$ mixing phase using $2.8\fb$ of data per experiment.
}

\FullConference{XXth Hadron Collider Physics Symposium \\
		 November 16 -- 20, 2009 \\
		 Evian, France}

\begin{document}

\section{Introduction}
In order to pursue new physics beyond the Standard Model (SM), 
the Tevatron $p\bar{p}$ collider with $\sqrt{s}=1.96\,\TeV$ provides powerful approach with $b$ hadrons.
At the Tevatron $b$ quarks are pair-produced with enormous cross section~\cite{Aaltonen:2009xn}, which is three orders of magnitude higher than at $e^+e^-$ colliders, and generate all sorts of $b$ hadrons.
This provides privileged access to SM-suppressed processes such as FCNC transitions and 
$CP$ violation in $\bs$ mixing.
These approaches from flavor sector at Tevatron are complementary to 
direct searches for BSM processes like Supersymetry (SUSY) particles,
and also $B$ physics at the $e^+e^-$ experiments.
In this paper we focus on studies for some promising FCNC processes; 
$B\to K^{(*)}\mu^+\mu^-$, $\bs\to \phi\mu^+\mu^-$,
$\bs\to\mu^+\mu^-$,
and
$\bs$ mixing,
performed by CDF and D0 collaborations.
\section{Rare decays}
\subsection{$B\to K^{(*)}\mu^+\mu^-$ and $\bs\to \phi\mu^+\mu^-$}
The $B\to K^{(*)}\mu^+\mu^-$ and $\bs\to \phi\mu^+\mu^-$ decays are dominated 
by the FCNC $\bsll$ transition.
In the SM framework, the quark transition is forbidden at the tree level.
It may occur via $Z$/$\gamma$ penguin diagram or a $W^+ W^-$ box diagram at the lowest order.
A new physics process could enhance the decay amplitude and it might be seen as an interference with the SM amplitude.
Therefore we measure various observables related to the magnitude or the complex phase, like branching ratio, polarization or forward-backward asymmetry.

\begin{figure}[b]
  \begin{center}
    \begin{tabular}{ccc}
      \resizebox{.25\textwidth}{!}{\includegraphics[clip]{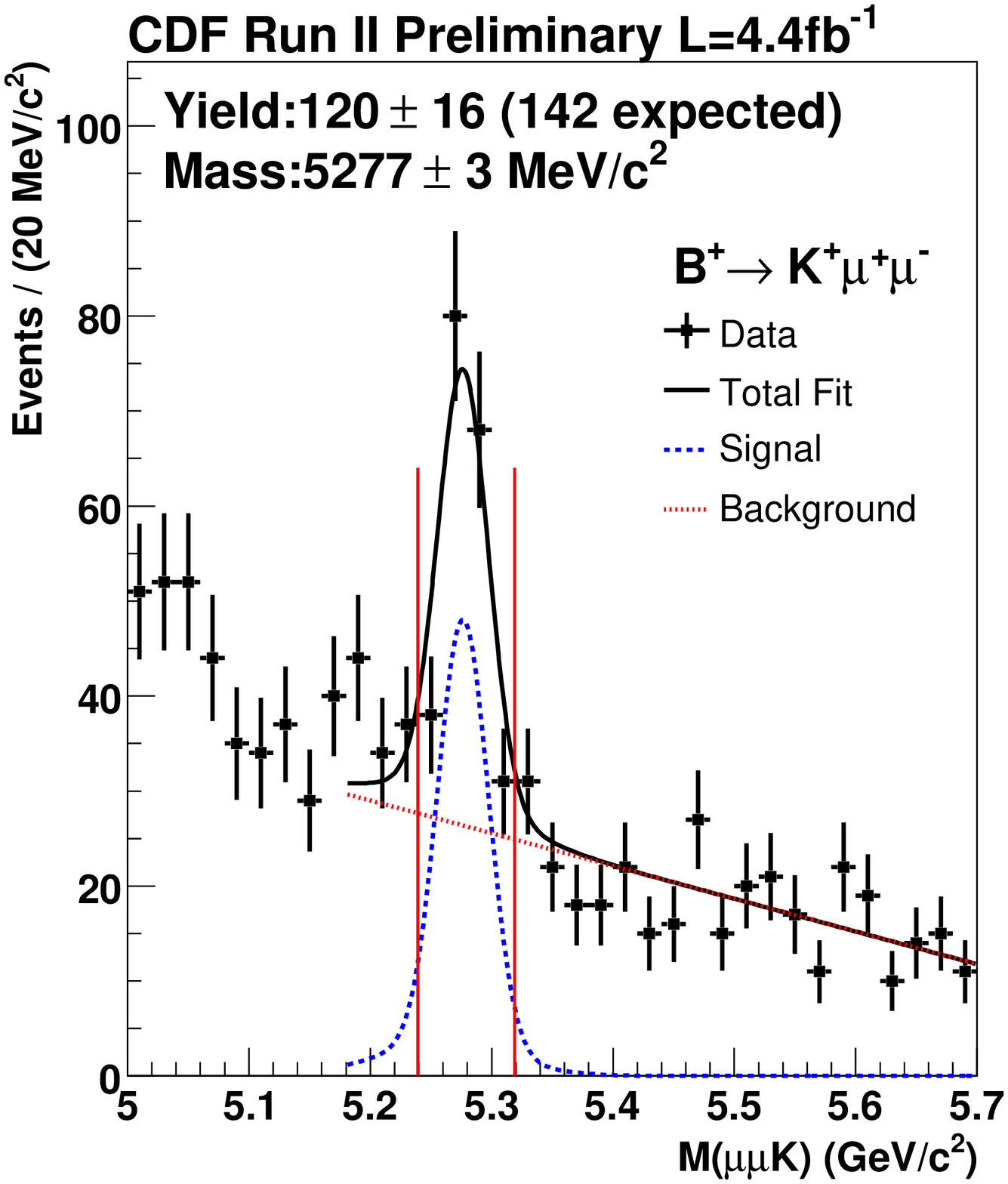}}
      \resizebox{.25\textwidth}{!}{\includegraphics[clip]{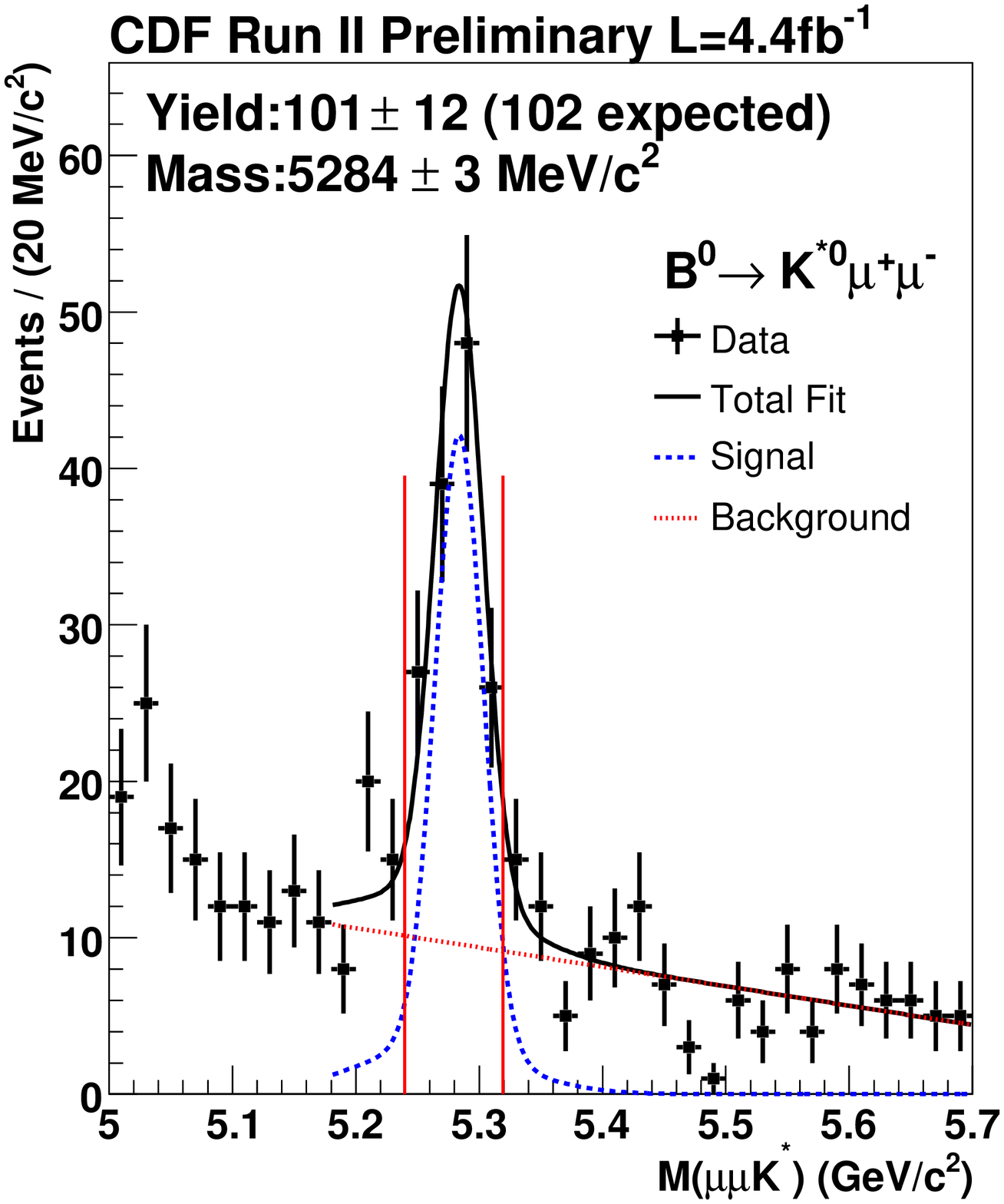}}
      \resizebox{.25\textwidth}{!}{\includegraphics[clip]{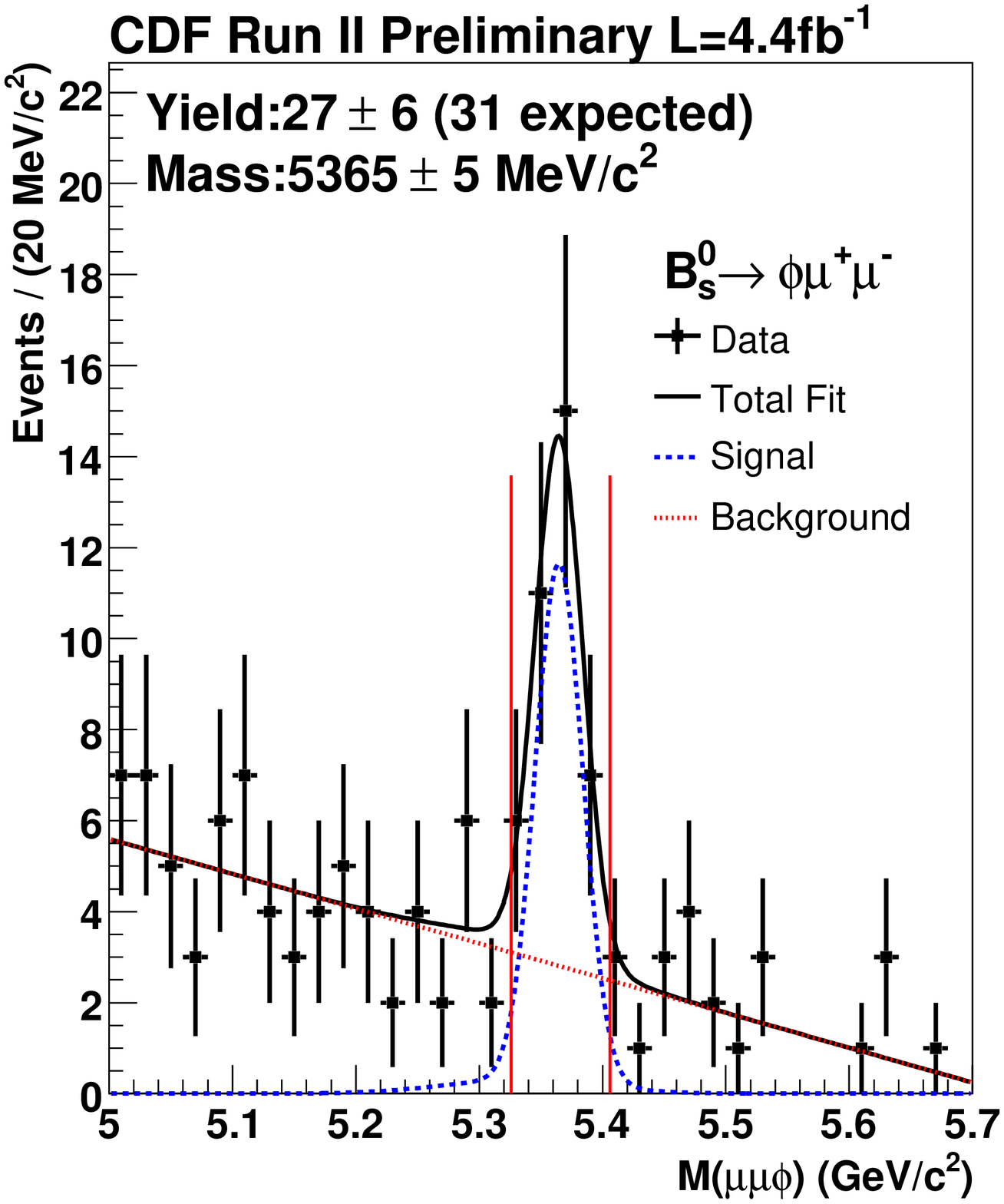}}
    \end{tabular}
  \end{center}
  \caption{The $B$ invariant mass of $\bpkmm$ (left), $\bzkstmm$ (middle), and $\bsphimm$ (right) 
for $\lumi$, respectively.
  }
  \label{fig:rarebmass_data}
\end{figure}

CDF selects two oppositely charged muon candidates with a momentum transverse to the beamline, 
$p_T$, greater than 1.5 or 2.0$\,\GeVc$, depending on the trigger selection.
We then reconstruct $\bhmm$ signal candidates, where $B$ stands for $\bp$, $\bz$, or $\bs$, and
$h$ stands for $K^+, \kst,$ or $\phi$ respectively.
The $\kst$ is reconstructed in the mode $\kst \to K^+\pi^-$, and 
the $\phi$ is reconstructed as $\phi \to K^+K^-$.
To enhance separation of signal from background we employ an artificial neural network (NN) technique.
Fig.~\ref{fig:rarebmass_data} shows invariant mass distribution for each rare decay.
The signal yield is obtained by an unbinned maximum log-likelihood fit of the $B$ invariant mass distribution.
From the $B$ mass fit with $4.4\fb$ of data~\cite{ref:cdf_10047}, 
we obtain $120\pm16$ ($\bpkmm$),  $101\pm12$ ($\bzkstmm$), and $27\pm6$ ($\bsphimm$) signal yields,
with $8.5\sigma$, $9.7\sigma$, and $6.3\sigma$ statistical significance, respectively.
This is the first observation of the $\bsphimm$ mode.
Obtained yields are consistent with world average and theoretical expectations.
We measure the branching fractions of rare decays relative to the corresponding reference channels, $\psih$,
which have same final states as rare decays but with an intermediate $\jpsi$ resonance.
Using PDG~\cite{Amsler:2008zzb} values for BR of reference decays we obtain
  ${\cal B}(\bpkmm) = [\kmmBr \pm \kmmBrStat({\rm stat}) \pm \kmmBrSyst({\rm syst})] \times 10^{-6}$,
  ${\cal B}(\bzkstmm) = [\kstmmBr \pm \kstmmBrStat({\rm stat}) \pm \kstmmBrSyst({\rm syst})] \times 10^{-6}$,
  ${\cal B}(\bsphimm) = [\phimmBr \pm \phimmBrStat({\rm stat}) \pm \phimmBrSyst({\rm syst})] \times 10^{-6}$.

\begin{figure}[t]
  \begin{center}
    \begin{tabular}{cc}
    \resizebox{0.33\textwidth}{!}{\includegraphics[clip]{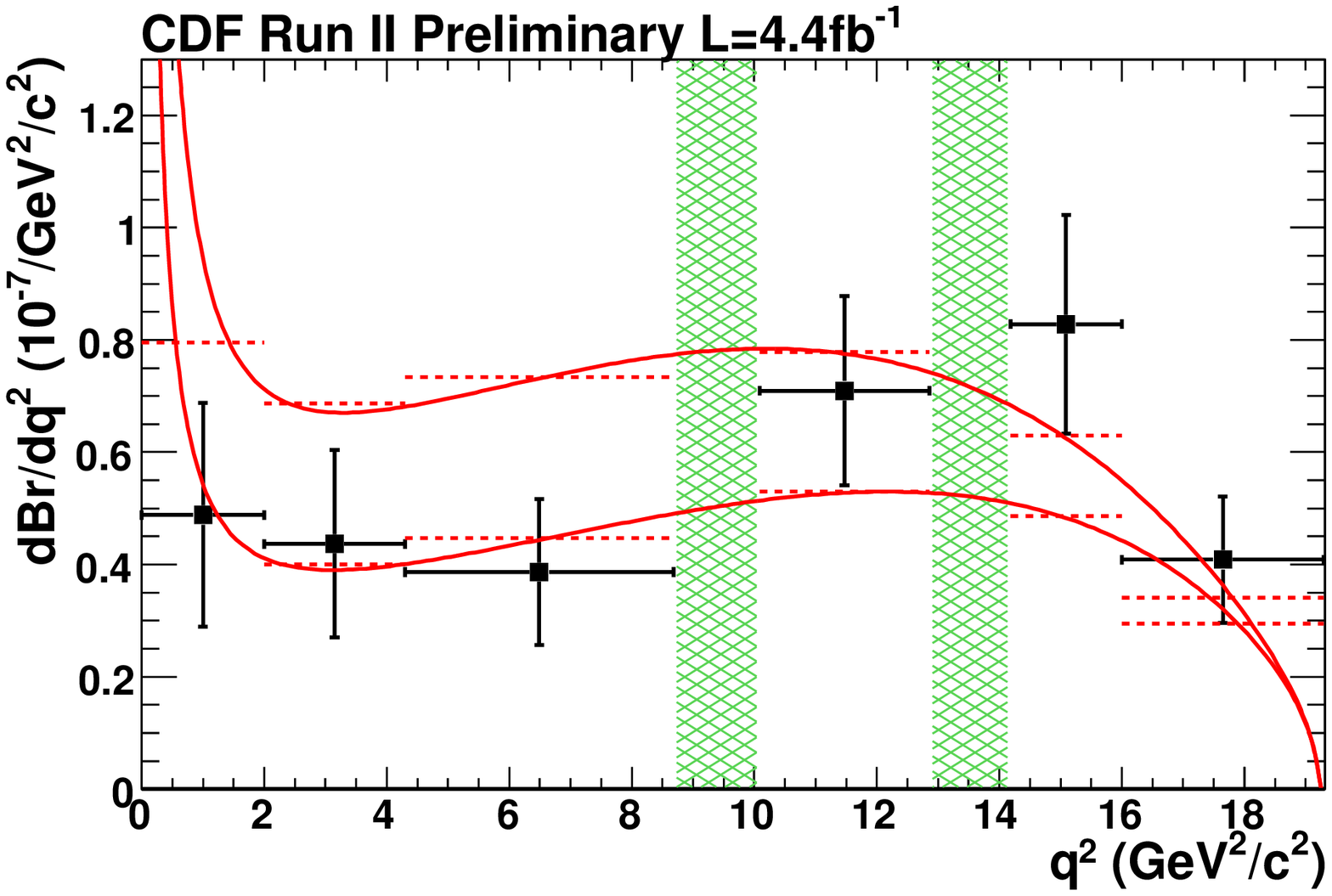}}
    \resizebox{0.33\textwidth}{!}{\includegraphics[clip]{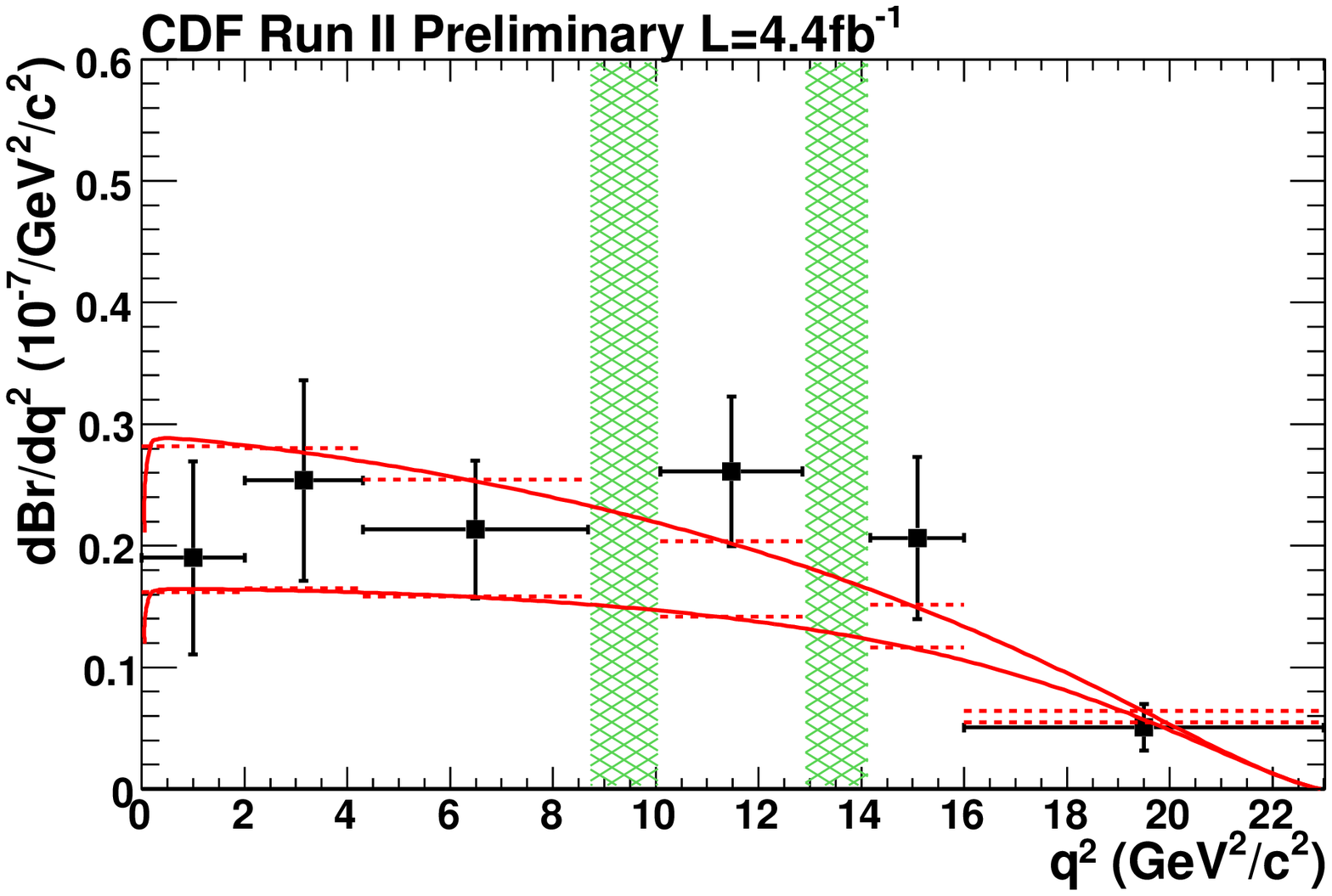}}
    \end{tabular}
  \end{center}
  \caption{Differential BR of $\bzkstmm$ (left) and $\bpkmm$ (right).
Hatched regions are charmonium veto regions.
 Solid lines are the SM expectation~\cite{Ali:1999mm}, which use maximum- and minimum- allowed form factor.
Dashed line is the averaged theoretical curve in each $q^2$ bin.
}
  \label{fig:dbr_kstmm}
\end{figure}

We measure the differential decay rate with respect to the dimuon mass.
The signal region is divided into six $q^2$ bins, where $q^2\equiv \Mmm c^2$.
To obtain the number of signal events in each $q^2$ bin we use 
the same procedure used in the global yield fit.
 Fig.~\ref{fig:dbr_kstmm} shows the differential branching fraction for $\bzkstmm$ and $\bpkmm$.

\begin{figure}[htypb]
  \begin{center}
    \begin{tabular}{ccc}
      \resizebox{.3\textwidth}{!}{\includegraphics[clip]{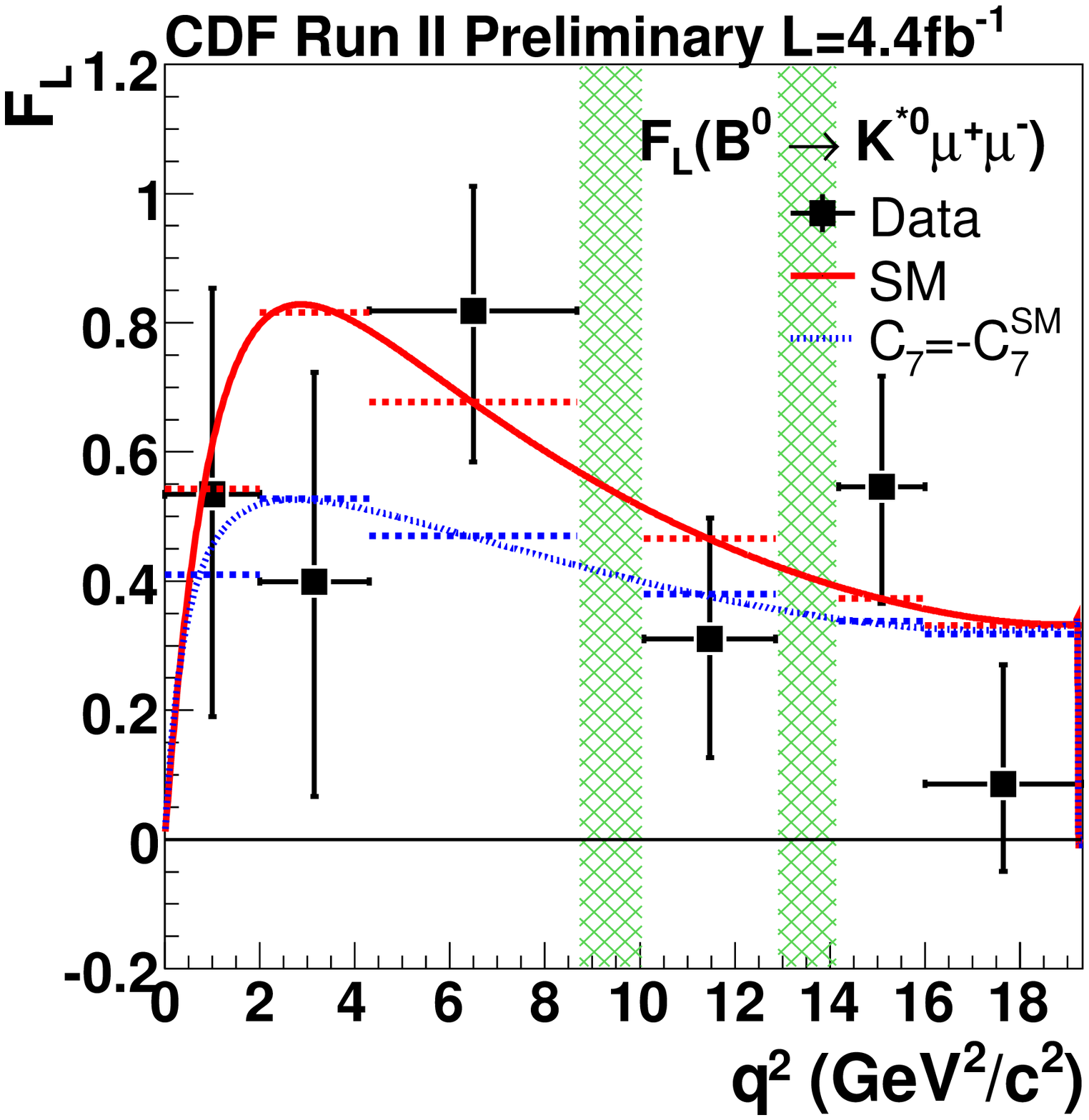}}
      \resizebox{.3\textwidth}{!}{\includegraphics[clip]{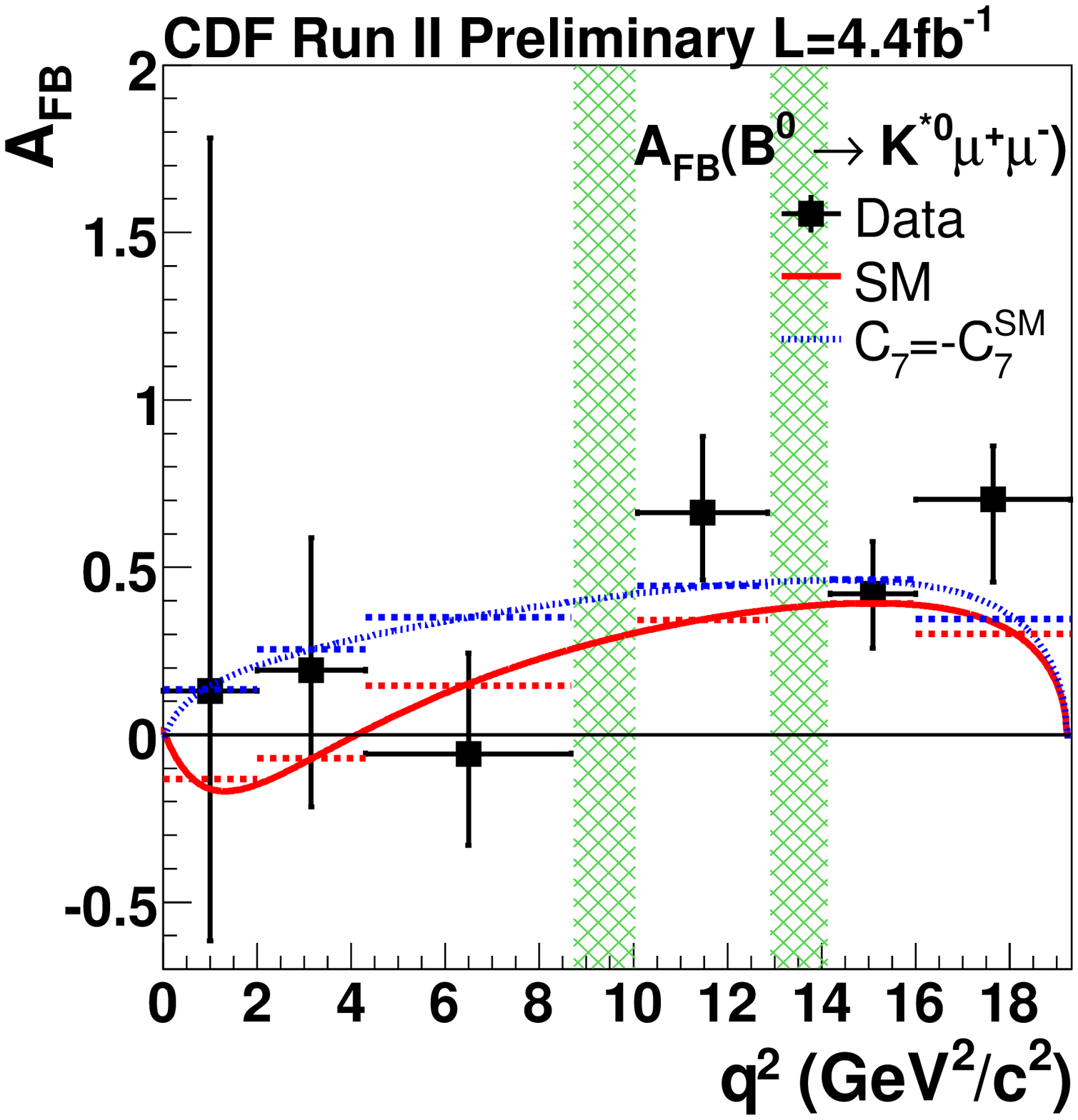}}
      \resizebox{.3\textwidth}{!}{\includegraphics[clip]{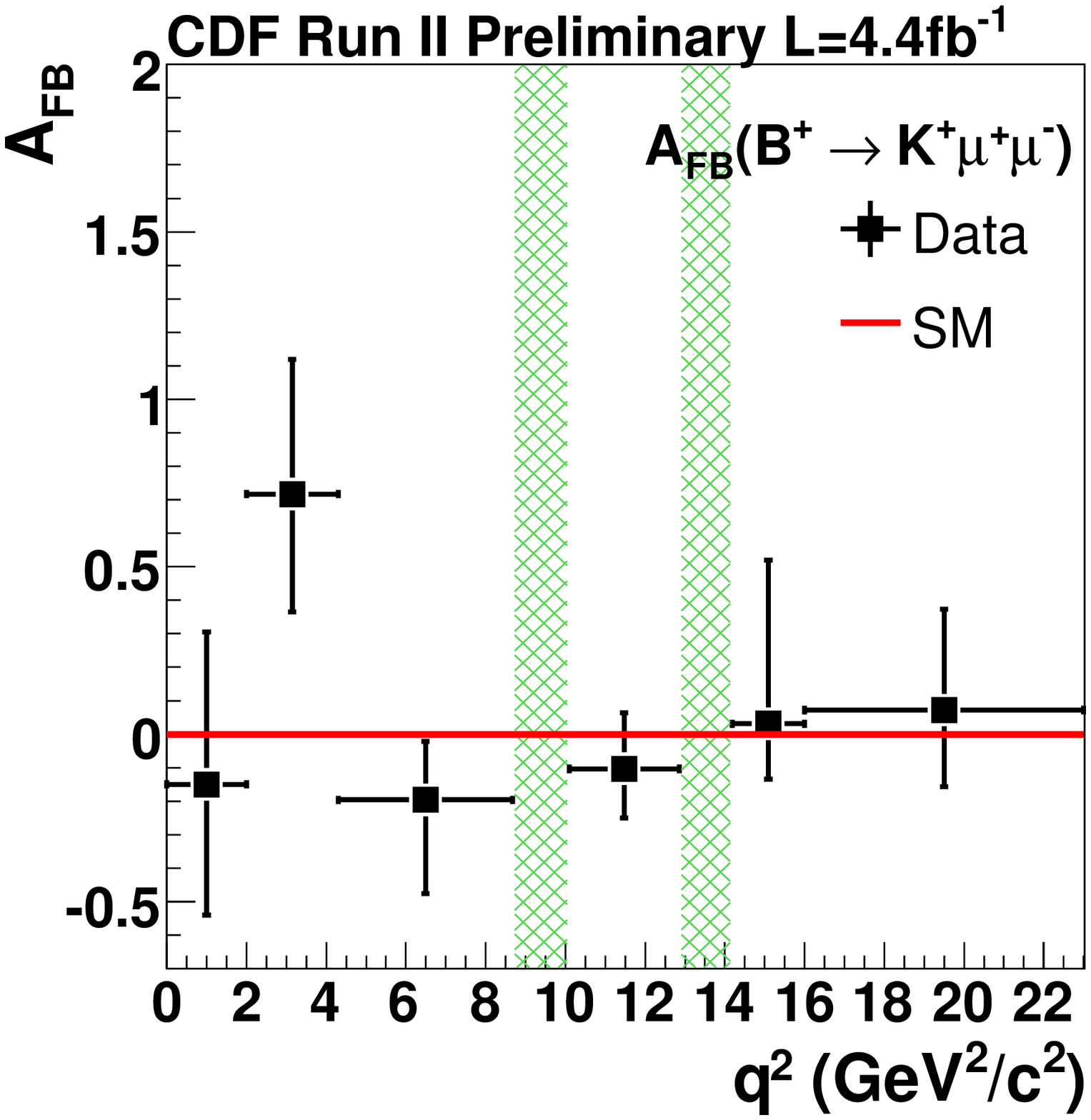}}
    \end{tabular}
  \end{center}
  \caption{$\fl$(left) and $\afb$(middle) fit results as a function of $q^2$ for $\bzkstmm$ and 
    $\afb$(right) as a function of $q^2$ for $\bpkmm$.
The points show data.
Solid (dotted) curve is the SM (an example of SUSY) expectation~\cite{Ali:1999mm}.
Dashed line is the averaged expectation in each $q^2$ bin.
Hatched regions mean charmonium veto.}
  \label{fig:afb_fl_kstmm}
\end{figure}

The forward-backward asymmetry ($\afb$) and $\kst$ longitudinal polarization ($\fl$) are extracted from $\cosm$ and $\cosk$ distributions, respectively, where
$\theta_\mu$ is the helicity angle between $\mu^+$ ($\mu^-$) direction and the opposite of the $B$ ($\overline{B}$) direction in the dimuon restframe, and
$\theta_K$ is the angle between the kaon direction and the direction opposite to the $B$ meson in the $\kst$ rest frame.
We measure $\fl$ and $\afb$ for $\bzkstmm$ and also $\afb$ for $\bpkmm$.
Fit results are shown in Fig.~\ref{fig:afb_fl_kstmm}.
Both $\fl$ and $\afb$ are consistent with the SM and also an example of SUSY model.
Our results are also consistent and competitive with B-factories measurements~\cite{Wei:2009zv,Aubert:2008ju}.

\subsection{$\bs (\bz) \to\mu^+\mu^-$}
The $\bs (\bz) \to \mu^+\mu^-$ decays 
are also dominated by FCNC process. The decay rates are
further suppressed by the helicity factor, $(m_\mu/m_B)^2$. 
The $\bz$ decay is also suppressed with respect to the $\bs$ decay
by the ratio of CKM elements, $\left|V_{td}/V_{ts}\right|^2$.
The SM expectations for these branching fractions are 
$\brBsmm = (3.42\pm0.54)\times10^{-9}$ and $\brBdmm = (1.00\pm0.14)\times10^{-10}$~\cite{Buras:2003td}.
As many new physics models can enhance the BR significantly, 
these decays provide sensitive probes for new physics.

\begin{figure}[htypb]
\begin{minipage}{0.5\hsize}
\begin{center}
\resizebox{0.7\textwidth}{!}{\includegraphics[clip]{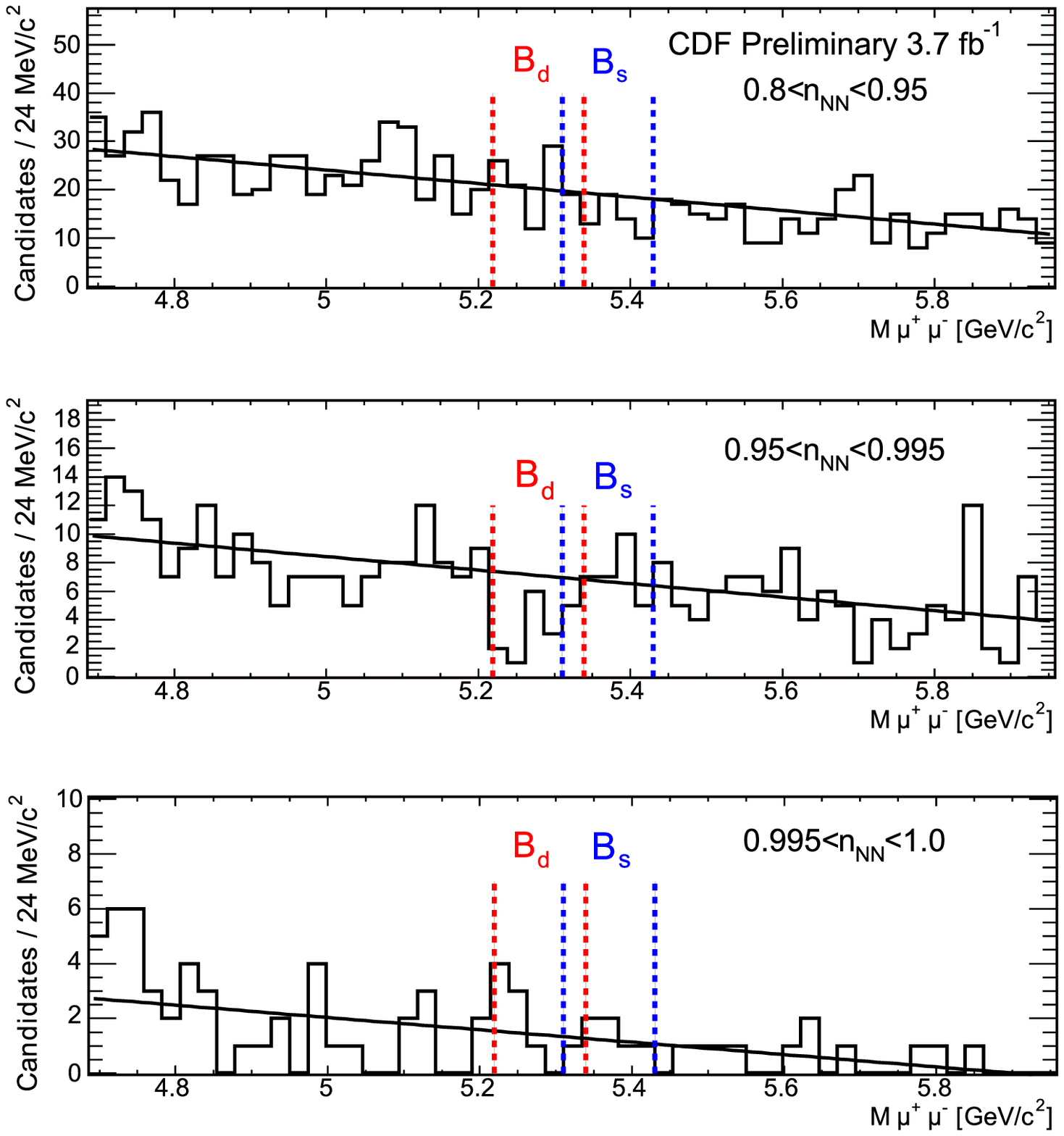}}
\end{center}
\end{minipage}
\begin{minipage}{0.5\hsize}
\begin{center}
\begin{tabular}{cc}
\resizebox{.6\textwidth}{!}{\includegraphics[clip]{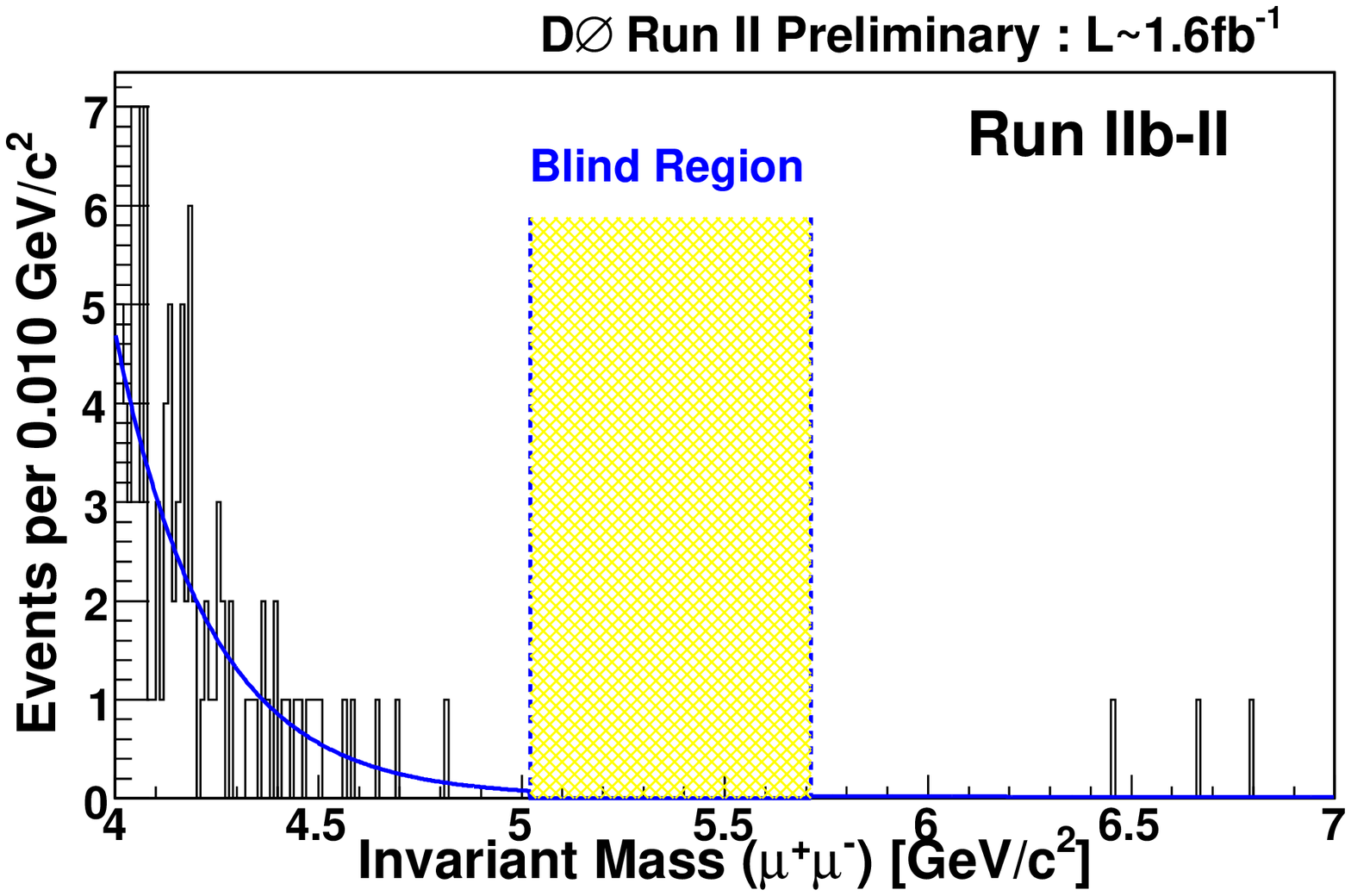}}\\
\resizebox{.6\textwidth}{!}{\includegraphics[clip]{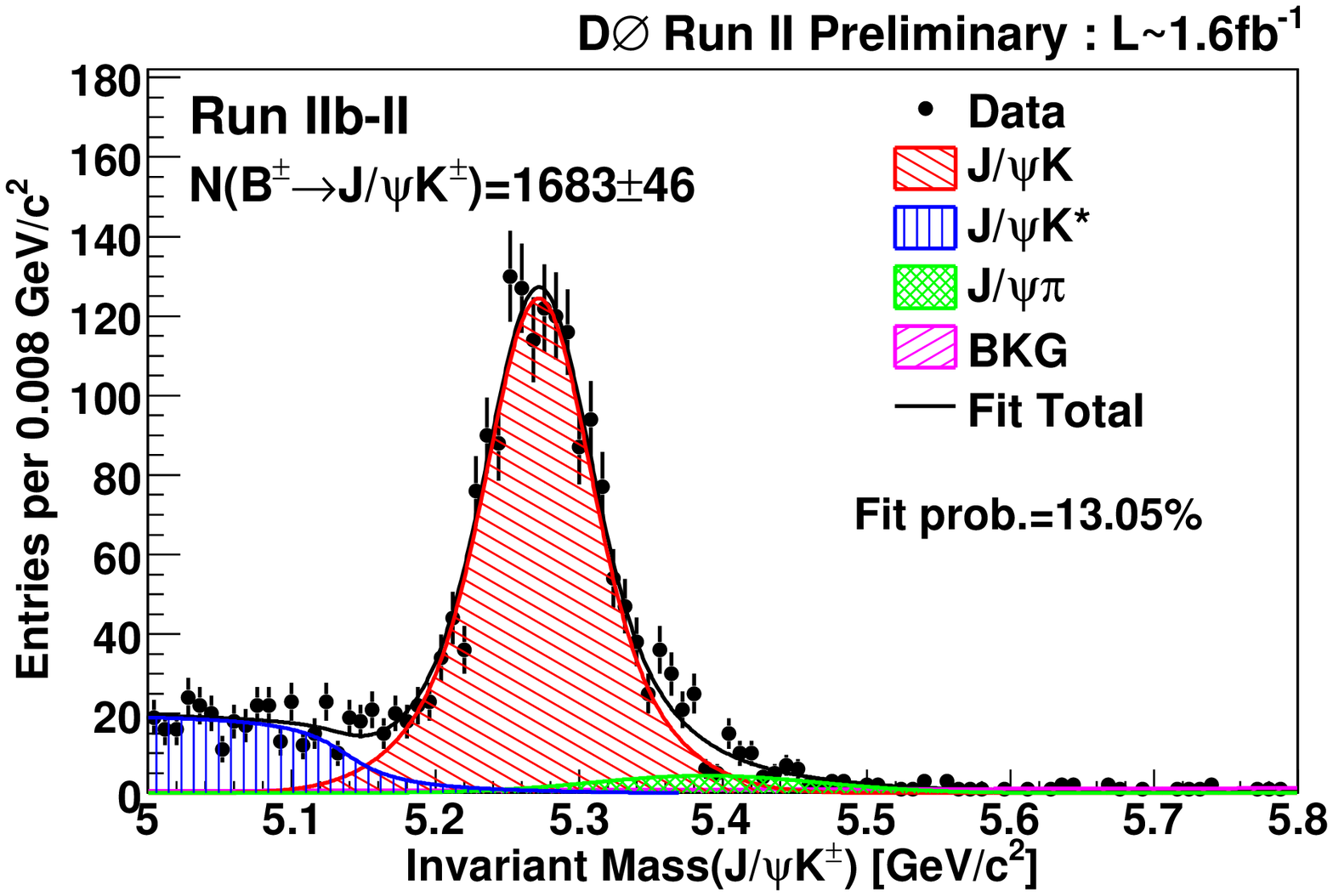}}
\end{tabular}
\end{center}
\end{minipage}
  \caption{Left: Dimuon invariant mass distribution for CDF events satisfying all selection criteria for the three ranges of NN.
Right top: Dimuon invariant mass distribution for D0 events after the BDT cut. Search box remains blinded.
Right bottom: $\jpsi K^+$ events used as a control sample, in D0 data after applying the BDT cut.
}
  \label{fig:bsmm}
\end{figure}

CDF selects two oppositely charged muon candidates within a dimuon invariant mass windows of
$4.669 < m_{\mu^+\mu^-} < 5.969 \,\GeVcsq$.
The muon candidates are required to have $p_T>2.0\,\GeVc$, and $\vec{p_T}^{\mu^+\mu^-}>4\,\GeVc$,
where $\vec{p_T}^{\mu^+\mu^-}$ is the transverse component of the sum of the muon momentum vectors.
For CDF analysis, we employ NN to select signal events.
The event selection is checked with 
control samples of $\bppsik$ and sideband in data that the background estimates are correctly predicted.
The $\mu^+\mu^-$ invariant mass distributions for the three different NN ranges are shown in Fig.~\ref{fig:bsmm},
using 3.7$\fb$ of data.
In the absence of signal, we extract 95\% (90\%) C.L. limits of $\brBsmm < 4.3\times 10^{-8}$ $(3.6\times 10^{-8})$~\cite{ref:cdf_9892} and
$\brBdmm < 7.6\times 10^{-8}$ $(6.0\times 10^{-8})$~\cite{ref:cdf_9892}, which are currently the world's best upper limits
for both processes.  

D0 performs a similar analysis but employs a Boosted Decision Tree (BDT) instead of NN.
With 5$\fb$ of data, D0 has studied the sensitivity to the 
branching fraction of $\Bsmm$ decays. An expected upper limit on the branching fraction is 
$\brBsmm\ < 5.3(4.3) \times 10^{-8}$ at the 95(90)\% C.L.~\cite{ref:d0_5906}.

\section{Measurement of the $\bs$ mixing phase}
Analogously to the neutral $\bz$ system, $CP$ violation in $\bs$ system may occur also through 
interference of decays with and without the $\bs$-$\bsb$ mixing.
The $\bs-\bsb$ mixing occurs via second order weak processes. It is described in the SM by $\Delta m_s$ and $\Delta \Gamma_s$, mass and decay width difference of the two mass eigenstates, $B_s^H$ and $B_s^L$.
The quantity $\Delta \Gamma_s=2|\Gamma_{12}|\cos(\phi_s)$ is sensitive to new physics effects
that affect the phase $\phi_s={\rm arg}(-M_{12}/\Gamma_{12})$,
where $\Gamma_{12}$ and $M_{12}$ are the off-diagonal elements of the mass and decay matrices.
In the SM, the $\phi_s^{\rm SM}$ is predicted to be small as 0.004~\cite{Lenz:2006hd}.
If new physics has different phase $\phi_s^{\rm NP}$ from the SM,
the $\phi_s$ could be dominated by $\phi_s^{\rm NP}$.
In this case we can access the phase by studying the time-evolution of $\bspsiphi$ decays.
The $CP$ violating phase $\beta_s^{\psiphi}$ is defined as the phase between the direct $\bspsiphi$ decay amplitude and mixing followed by decay amplitude.
The $\beta_s^{\rm SM}$ is described by CKM matrix elements as ${\rm arg}(-V_{ts}V_{tb}^*/V_{cs}V_{cb}^*)$ and predicted to be small, 0.02~\cite{Lenz:2006hd}.
Since $\phi_s^{\rm NP}$ contributes to both $\phi_s$ and $\beta_s$,
large $\beta_s$ would indicate existence of new physics contribution.

To extract $\Delta \Gamma_s$ and $\beta_s$, an unbinned maximum likelihood is performed.
The $\bspsiphi$ consists of both $CP$-even and -odd final states. 
Although the observed $CP$ asymmetry might be diluted by the opposite $CP$ components,
we can perform unbiased measurement taking into account time evolution of angular distributions of the decay products.
Information about mixing is obtained from flavor tagging of $\bs$ meson, which is based on
kaon tracks associated with the $\bs$ meson and
the properties and decay tracks of the other $B$ hadron in the event.
Since there is an exact symmetry in the signal probability density function, which contains the strong phases among the three partial waves,
the likelihood function shows two symmetric minima in the $\Delta \Gamma_s$-$\beta_s^{\psiphi}$ plane.

\begin{figure}[htypb]
  \begin{center}
    \begin{tabular}{cc}
      \resizebox{.39\textwidth}{!}{\includegraphics[clip]{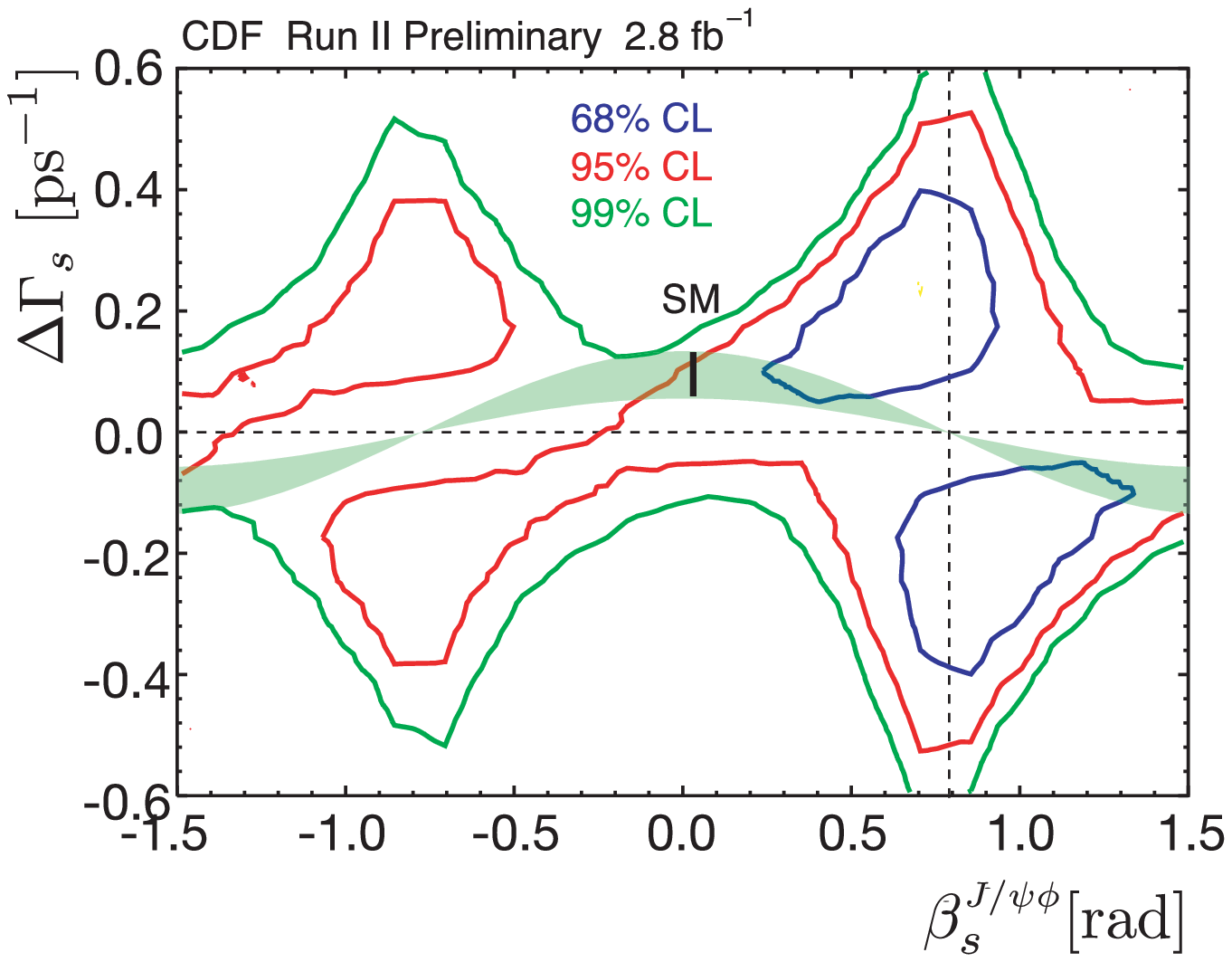}} & 
      \resizebox{.39\textwidth}{!}{\includegraphics[trim = 340 0 0 0 , clip]{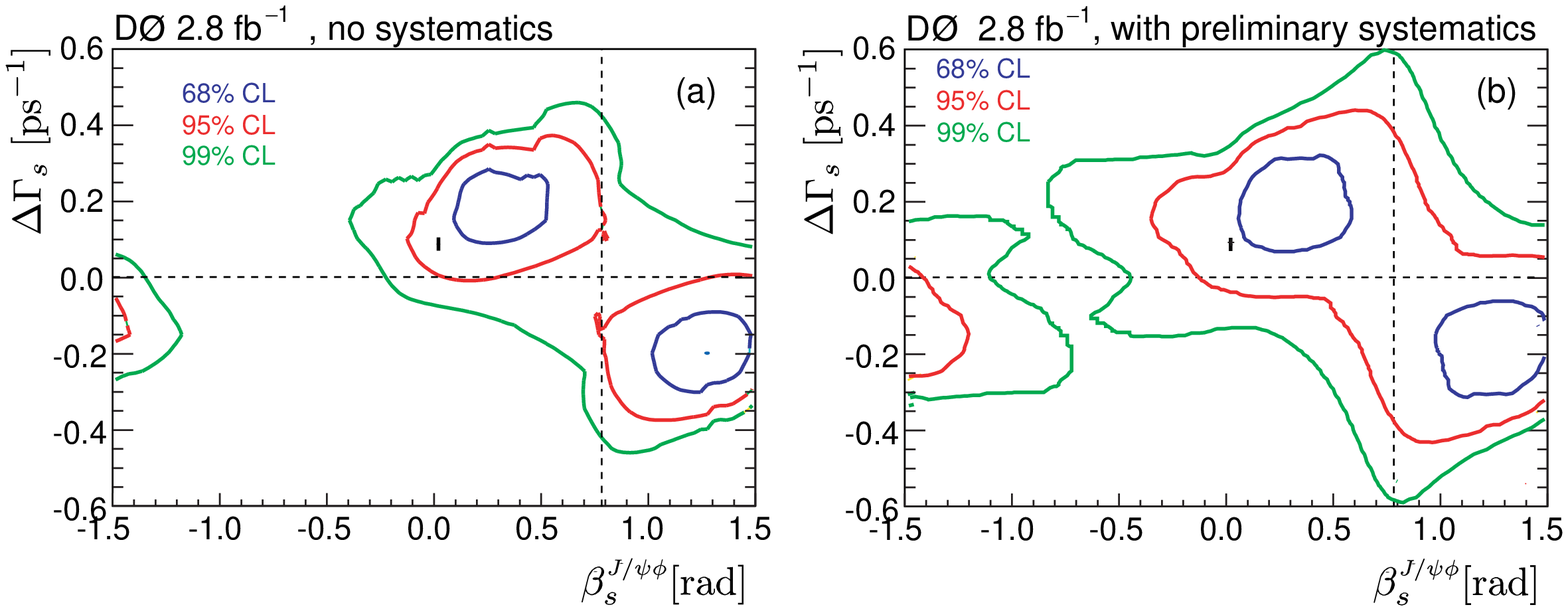}}\\
      \resizebox{.39\textwidth}{!}{\includegraphics[clip]{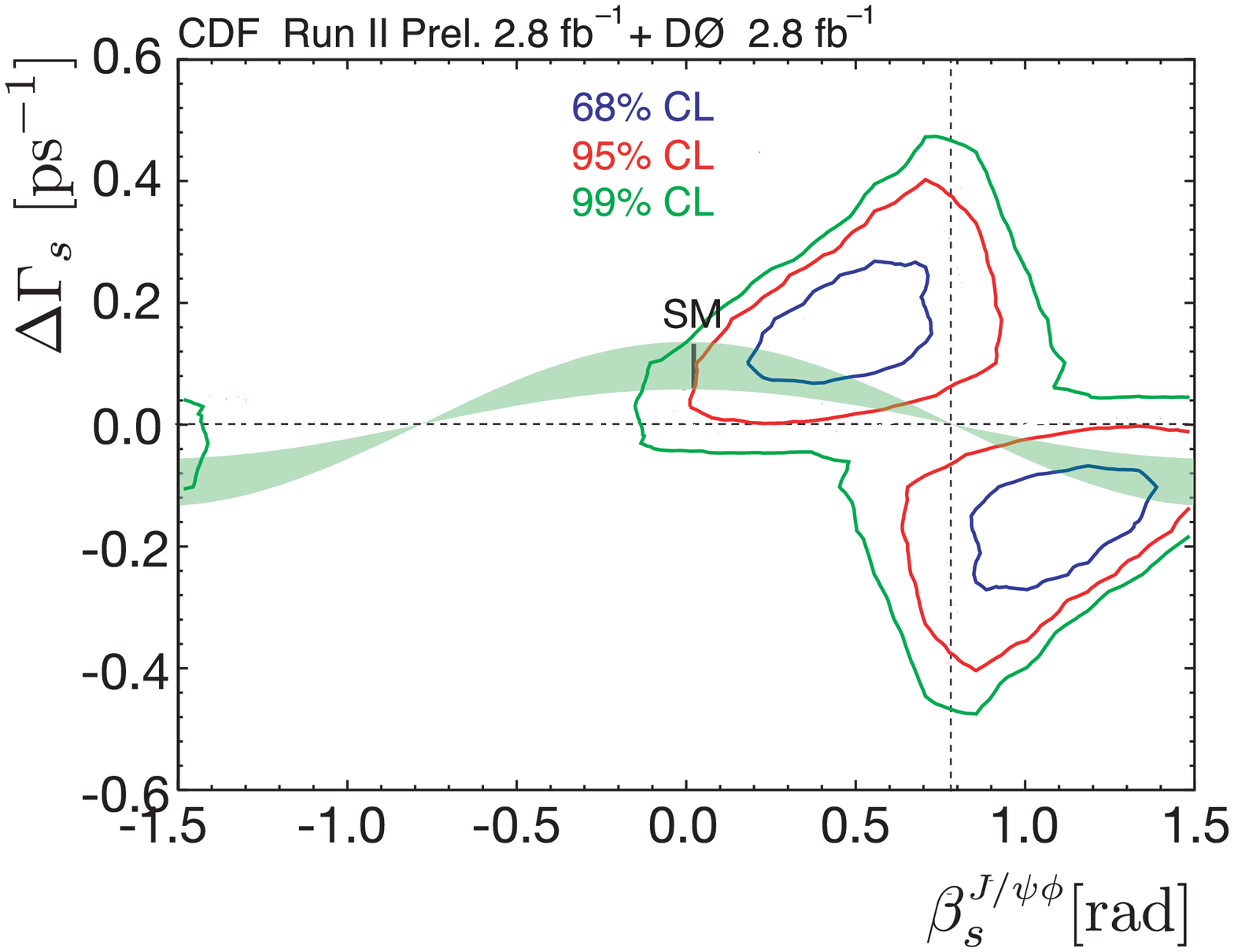}} & {} \\
      \end{tabular}
  \end{center}
  \caption{
Two-dimensional profile likelihood as confidence contours of $\beta_s^{\jpsi\phi}$ and $\Delta\Gamma_{s}$ for CDF's preliminary analysis using $2.8\fb$ of data (top left) and
D0's published analysis using $2.8\fb$ of data,
but allowing strong phases, $\delta_i$ to float and systematic uncertainties are included (top right),
and the combined results (bottom).
The SM expectation and uncertainty ($\beta_s^{\rm SM}$, $\Delta\Gamma_{s}^{\rm SM}$)=
($0.04$, $0.088\pm0.017{\rm ps}^{-1}$)~\cite{Lenz:2006hd} is indicated by the black line.
 The region allowed in new physics model given by $\Delta\Gamma_{s}=2|\Gamma_{12}|\cos\phi_s$ 
 is also shown (light green band).
}
  \label{fig:betas_cdf_d0}
\end{figure}

Both CDF and D0 have performed flavour tagged analysis on 2.8$\fb$ of data~\cite{ref:cdf_9458,Abazov:2008fj}.
CDF selected about 3200 signal events with NN, while D0 selected about 2000 signal events with a cut based selection.
Fig.~\ref{fig:betas_cdf_d0} (top left) shows the confidence regions for CDF and
Fig.~\ref{fig:betas_cdf_d0} (top right) shows the fit result for D0.
D0 updates the result from their previous publication result, which restricted the strong phases $\delta_{||}$ and $\delta_{\perp}$ to the values measured in the $\bz \to \jpsi K^{*0}$ system.
D0 removes the constraints and also includes systematic uncertainties on $\Delta m_s$.
Currently the compatibility with the SM point is 1.8$\sigma$ for CDF and 1.2$\sigma$ for D0.
We then combine both profile likelihoods. Detail of combination is described in Ref.~\cite{ref:cdf_9787}.
Fig.~\ref{fig:betas_cdf_d0} (bottom) shows the combined results of CDF and D0, which exhibit a $2.1\sigma$ deviation from the SM.

\section{Conclusion}
At the Tevatron a rich $B$ physics program is ongoing.
CDF reports the first observation of $\bsphimm$ and measures $\afb(B\to K^{(*)}\mu^+\mu^-)$ in hadron collisions,
which is competitive to $e^+e^-$ B-factories.
CDF updates the $\bs(\bz)\to \mu^+\mu^-$ analysis using $3.7\fb$ and continues to improve its world-leading upper limit.
D0 continues to improve their $\bs\to \mu^+\mu^-$ analysis and with $5\fb$ their expected limit is 
$\brBsmm\ < 5.3(4.3) \times 10^{-8}$ at the 95(90)\% C.L..
Both CDF and D0 have updated their $\beta_s^{\psiphi}$ measurements with $\bspsiphi$ using $2.8\fb$ of data.
Combined result of both experiments shows a $2.1\sigma$ deviation from the SM.
The Tevatron is performing well with planed running through 2011 will provide double the datasets used for results presented here.


\end{document}